\documentclass[12pt]{article}
\pdfoutput=1
\usepackage{amsmath, amssymb}    
\usepackage{graphicx}% Include figure files
\usepackage{dcolumn}% Align table columns on decimal point
\usepackage{bm}% bold math
\usepackage{hyperref}
\usepackage{color, xcolor}
\usepackage{tikz}
\usepackage{tikz-feynman}

\newcommand{\be}{\begin{equation} } 
\newcommand{\ee}{\end{equation} } 
\newcommand{\ba}{\begin{array} } 
\newcommand{\ea}{\end{array} } 
\newcommand{\bear}{\begin{eqnarray} } 
\newcommand{\eear}{\end{eqnarray} } 
\newcommand{\met}{ E_T\!\!\! \!\! \! \slash \;\, }

\begin{document}

\baselineskip=18pt \pagestyle{plain} \setcounter{page}{1}

\vspace*{-1cm}

\noindent \makebox[11.9cm][l]{\small \hspace*{-.2cm} }{\small Fermilab-PUB-26-0730-T}  \\  [-1mm]

\begin{center}

{\normalsize \large  \bf  \hspace*{-.3cm}  
Minimal cascade decay of the top quark through a new scalar  
\\ [9mm]
}
\vspace*{0.3cm}

{\bf  Bogdan A. Dobrescu$^\diamond$, Max H. Fieg$^\diamond$, and Alessandro Russo$^\star$} 
\vspace{5mm}
\\
{\normalsize\small 
\it
$^\diamond$  Particle Theory Group, Fermilab, Batavia, IL 60510     \\[3.5mm]
$^\star$  SLAC National Accelerator Laboratory, Menlo Park, CA 94025 
 \\[2mm]
$^\star$ Leinweber Institute for Theoretical Physics, Department of Physics, \\ Stanford University, Stanford, CA 94305 
\let\thefootnote\relax
\footnotetext[0]{bdob@fnal.gov, mfieg@fnal.gov, arusso00@stanford.edu}
}

\vspace{9mm}

{ \normalsize\small  October 5, 2026 }

\vspace{9mm}

\begin{abstract}  
We explore a possible decay of the top quark into a new scalar $\phi_t$ plus a charm quark. The decay is mediated by a dimension-5 operator that may be generated by the exchange of a heavy vectorlike quark, $\chi$, and can have a branching fraction close to a percent. In a minimal model, $\phi_t$ decays due to the same operator that allows the nonstandard top decay, leading to a final state involving a $c$ jet, a $b$ jet and  a $W$ boson. Hence, the top quark would undergo a nonstandard cascade decay through $\phi_t$ and then through an off-shell top quark. The main nonstandard collider signal is a $t \bar t \to c \, \bar c \, b \, \bar b \, W^+ W^-$  process. In addition, production of the vectorlike quark at the LHC would lead to a signal of the type $\chi \phi_t \, j \to  3c + b \, \bar b \, W W j$.  
\end{abstract} 

\end{center}

\vfil

\thispagestyle{empty}  
  
\setcounter{page}{1}  
  
\vspace*{0.31cm}    
  
\newpage   
  
\tableofcontents
  
\vspace*{0.1cm}    
  
\baselineskip18pt   

\bigskip\bigskip

%%%%%%%%%%%%%%%%%%%%%%%%%%%%%%%%%%%%%%%%%%%%%%%%%%%%%%%%%%%%%%%%%%%%%%%%%%%%
\section{Introduction} 
\label{sec:intro}

In the 31 years that have passed since the discovery of the top quark by the CDF \cite{CDF:1995wbb} and D0 \cite{D0:1995jca} experiments at the Tevatron, more than $4\times 10^8$ $t\overline t$  pairs have been produced at hadron colliders. The majority of $t\overline t$  pairs have been produced since 2016 at the LHC within the ATLAS and CMS experiments, which have searched for various possible decay modes of the top quark into Standard Model (SM) particles. Despite the large number of top quarks produced and a broad set of searches (\cite{ATLAS:2024mih, CMS:2021gfa, ATLAS:2024njy, CMS:2023iul, CMS:2024dzv, ATLAS:2023qzr} and many other searches referenced in \cite{ParticleDataGroup:2026mpi}), only one decay mode has ever been observed, namely the SM tree-level one, $t \to W^+ b$. 
This is peculiar given that all known elementary particles heavier than the muon have many measured decay modes \cite{ParticleDataGroup:2026mpi}, and furthermore the top quark is the heaviest elementary particle discovered so far.

In this article we investigate the possibility that the top quark has a second measurable decay mode, into a charm quark and a new spin-0 particle. Specifically, we consider a gauge-singlet complex scalar $\phi_t$ that participates in a dimension-5 operator involving the SM third generation quark doublet, the weak-singlet charm quark, and the SM Higgs doublet. That operator is induced in a simple renormalizable theory by the exchange of a vectorlike quark $\chi$ that has the same SM gauge charges as the weak-singlet top quark.
A vectorlike quark of this type has been identified as an essential ingredient in composite Higgs models \cite{Dobrescu:1997nm,Chivukula:1998wd}, and uncovering even indirect effects of it would open a window into the underlying new physics at the 10 TeV scale.

We will argue that the mass of $\chi$ is constrained by some CMS \cite{CMS:2022fck} and ATLAS \cite{ATLAS:2024gyc} searches to be above roughly 1.5 TeV. 
This lower mass limit, in conjunction with the  upper limit (of order one) on the Yukawa couplings of the vectorlike quark, implies that the branching fraction of the nonstandard decay of the top quark, $t \to \phi_t \, c$ is below percent level. Even though  this may seem like a small branching fraction, it could have led to millions of nonstandard $t \overline t$ events already produced at the LHC. 

The simplest hypothesis for the decay modes of $\phi_t$, is that the same dimension-5 operator responsible to the nonstandard top decay is the only Lagrangian term that can induce a $\phi_t$ decay. The leading process of this type for a $\phi_t$ mass in the $\sim 90 - 170$ GeV range is a 3-body decay: $\phi_t \to \overline c \, t^* \to \overline c \, b \, W^+$, where the $*$ indicates an  off-shell particle. 
Thus, the top quark undergoes a cascade decay  that involves an off-shell top quark: 
\be
t \to c\, \phi_t \to c \overline c \, t^* \to c \overline c \, b \, W^+  ~~.
\label{eq:topdecay}
\ee
We will discuss the special features of this nonstandard top decay in the case of $t\overline t$ production, and argue that the CMS and ATLAS Collaborations would be sensitive to the main signal, $c \overline c \, b \overline  b \, W^+W^-$, in a significant region of parameter space.

It is essential that studies of nonstandard processes based on higher-dimensional operators are correlated with plausible origins of the operators within renormalizable theories. Otherwise, the scale that suppresses those operators may be assumed to lie within ranges that are already ruled out by other searches. In the case of nonstandard top processes, the heavy particles that are integrated out may be directly produced at the LHC, and there is competition between searches for them and the collider searches for new top decay modes.

Within the minimal model studied here, the vectorlike quark has an interesting decay mode, $\chi \to \phi_t \, c$, which is dominant for a range of Yukawa couplings.  Based on that decay,  single production of the vectorlike quark in association with a $\phi_t$ leads to a $3c + b \, \bar b \, W W j$ final state. 
We will identify some features of this new signal that can be targeted by the ATLAS and CMS experiments, and will discuss the complementarity between searches for the vectorlike quark and searches for the nonstandard top decay (\ref{eq:topdecay}). 

Top quark decays through a new spin-0 particle have been the subject of several previous studies.
Bahl, Koren and Wang \cite{Bahl:2023xkw} analyzed various higher-dimensional operators (including the one analyzed here), calculated the partial width for the top decay to a lighter quark and a new scalar, and also for the tree-level and loop-induced decay modes of the scalar.
The decay to a scalar $S$ and a charm quark, $t \to S c$, followed by $S \to b \bar b$ or $\gamma\gamma$ was studied in \cite{Banerjee:2018fsx},  while that decay followed by $S \to \mu \tau$ was studied in \cite{Bhattacharyya:2022umc}.
Displaced vertices from top-charm interactions of axion-like particles were considered in \cite{Cheung:2024qve, Carmona:2022jid}, and have been searched for by the ATLAS Collaboration  \cite{ATLAS:2024grv}. 
We focus here on prompt signals of the complex scalar produced in the top decay. A search of this type, but in a different final state, 
was reported in \cite{ATLAS:2023mcc}.  

In Section \ref{sec:model} we extend the SM with a complex scalar that has an effective interaction with the top and charm quarks, which is generated when the vectorlike quark is integrated out. 
A detailed computation of the complex scalar width is performed in Section~\ref{sec:3body}, and is followed by a discussion of the LHC signals of the vectorlike quark.
In Section \ref{sec:LHC} we analyze  the nonstandard top-quark decay at the LHC. Our conclusions are summarized in Section \ref{sec:conclusions}.

\medskip \medskip \medskip  

%%%%%%%%%%%%%%%%%%%%%%%%%%%%%%%%%%%%%%%%%%%%%%%%%%%%%%%%%%%%
%%%%%%%%%%%%%%%%%%%%%%%%%%%%%%%%%%%%%%%%%%%%%%%%%%%%%%%%%%%%
\section{Complex scalar coupled to top and charm quarks}     
\label{sec:model}   \setcounter{equation}{0}

We consider an extension of the SM that includes a gauge-singlet complex scalar $\phi_t$ that has no VEV and experiences  
only one interaction with the SM fields:
\be
\frac{1}{\Lambda_5} \phi_t  H \,  \overline q^3_{_L}  c_{_R}  + {\rm H.c.}
\label{eq:op5}
\ee
Here $q_{_L}^3 = (t_{_L}, b_{_L})$ is the third-generation SM quark doublet, $c_{_R}$ is the right-handed  charm quark,  $H$ is the SM Higgs doublet, and $\Lambda_5$ is the scale of this dimension-5 operator. 
As we will show below, it is self-consistent to assume that the above operator is the only one involving $\phi_t$. 
A similar operator, $\phi_t  H \,  \overline q^2_{_L}  t_{_R} $ where $q_L^2 = (c_{_L}, s_{_L})$ is the second-generation SM quark doublet and $t_R$ is the right-handed SM top quark, would also be interesting (see \cite{Bahl:2023xkw} for a discussion of that operator), but will be ignored here.  

Replacing $H$ by its VEV in (\ref{eq:op5}) yields an effective Yukawa interaction 
\be
\frac{v_{\! _H}  }{\Lambda_5} \, \phi_t   \, \overline t_{_L}  c_{_R}  + {\rm H.c.} ~,
\label{eq:phitc}
\ee
where the normalization of the weak scale is $v_{\! _H} \approx 174$ GeV.
This interaction allows the nonstandard top quark decay $t \to \phi_t \, c$ for a $\phi_t$ mass $M_\phi \lesssim 170$  GeV.
The tree-level width for that 2-body decay, 
for $M_\phi$ satisfying $\epsilon_c \equiv m_c^2 / (m_t - M_\phi)^{2} \ll  1$ with $m_c$ the charm quark mass, is given by
\be
\Gamma (t \to \phi_t \, c) = \frac{v_{\! _H}^2}{32 \pi \Lambda_5^2}  \, m_t   \left( 1 -  \frac{M_\phi^2}{m_t^2} \right)^{\! \! 2} \,
\left( 1 - \frac{2 M_\phi^2  \,  m_c^2}{(m_t^2 - M_\phi^2)^2} + O\!\left(\epsilon_c^2\right)  \right) ~~.
\label{eq:width}
\ee

This needs to be compared with the SM tree-level top width,
\be
\Gamma (t \to W^+ b) = \frac{g_w^2 \, m_t^3  }{64 \pi M_W^2}  \left( 1 -  \frac{M_W^2}{m_t^2} \right)^{\! 2}  \left( 1 + 2\frac{M_W^2}{m_t^2} \right) ~~,
\label{eq:SMtwidth}
\ee
where $g_w \approx 0.65$ is the $SU(2)_W$ gauge coupling at the $m_t \approx 172.6$ GeV scale \cite{ParticleDataGroup:2026mpi}, and the small corrections due to the $b$ quark mass (of order $m_b^2/m_t^2$) are ignored.
The QCD corrections to the $t \to W^+ b$ width decrease the tree-level result by approximately 10\% \cite{Li:1990qf,Beneke:2000hk}, and would change sign if the $M_W/m_t$ ratio were closer to 1.

As the matrix element for  the $t \to \phi_t \, c$ width involves a scalar instead of a gauge boson, the QCD corrections are likely to have a different behavior than the SM one. To compute the NLO QCD corrections, we first generated UFO \cite{Darme:2023jdn} model files based on the Lagrangian terms of the complex scalar $\phi_t$ included in FeynRules \cite{Alloul:2013bka},
which relies on loop computations provided by NLOCT \cite{Degrande:2014vpa} and FeynArts \cite{Hahn:2000kx}. Those model files are then used in simulations by \texttt{MadGraph} \cite{Alwall:2014hca}, which involves several computer packages: MadLoop \cite{Hirschi:2011pa}, MadFKS \cite{Frederix:2009yq}, Ninja \cite{Hirschi:2016mdz}, CutTools \cite{Ossola:2007ax}. 

Before computing the nonstandard branching fraction of the top, we need to estimate the lower limit on the scale $\Lambda_5$ of the 
dimension-5 operator (\ref{eq:op5}). To that end, it is important to consider possible renormalizable origins of that operator. 
A renormalizable theory of that type includes a vectorlike quark, $\chi$, transforming as $(3, 1, +2/3)$ under  the $SU(3)_c \times SU(2)_W \times U(1)_Y$ gauge group of the SM. Its Yukawa interactions with $\phi_t$ and $H$ are given by 
\be
- y_{\chi \phi} \, \phi_t \, \overline \chi_{_L} c_{_R}  -  y_{\chi \hspace*{-0.3mm} { \scriptscriptstyle H} } \, H \,  \overline  q^3_{_L} \chi_{_R}  + {\rm H.c.} 
\label{eq:Yukawa}
\ee 
The Yukawa couplings $y_{\chi \phi}$ and $y_{\chi \hspace*{-0.3mm} { \scriptscriptstyle H} }$ can be taken positive without loss of generality (their phases can be absorbed in the  $\phi_t$ and $\chi$ fields),  and are generically of order one or smaller. 
The vectorlike quark has a mass term $- m_{\chi} \, \overline \chi \chi$, with $m_{\chi}  \gg m_t$. 
Integrating out $\chi$ at tree level gives operator (\ref{eq:op5}) with a scale related to the $\chi$ couplings and mass: 
\be
\Lambda_5 = \frac{m_{\chi}}{y_{\chi \phi} \,  y_{\chi \hspace*{-0.3mm} { \scriptscriptstyle H} }}  ~~.
\ee

%%%%%%%%
\begin{figure}[t!]
\centerline{\hspace*{-.6cm}   \includegraphics[width=10.5cm, angle=0]{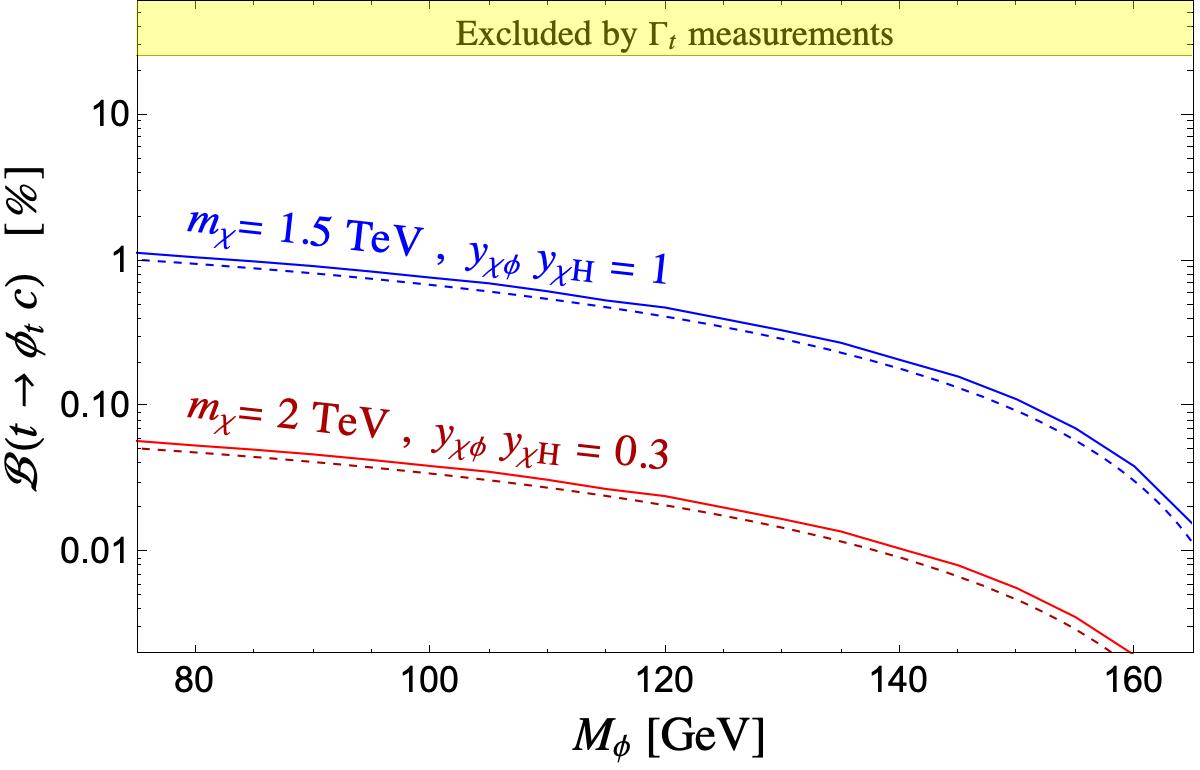}    }
\vspace*{0.cm}  
\caption{Nonstandard branching fraction of the top quark decay into a complex scalar $\phi_t$ and a charm quark, as a function of the $\phi_t$ mass. The decay proceeds through the dimension-5 operator (\ref{eq:op5}) induced by a vectorlike quark of mass  $m_\chi$, taken to be 1.5 TeV (upper blue lines) or 2 TeV (lower red lines). 
The product of the Yukawa couplings defined in (\ref{eq:Yukawa}) is taken here to be $y_{\chi \phi} y_{\chi \hspace*{-0.3mm} { \scriptscriptstyle H} } = 1$ (0.3) for the blue (red) lines.
The solid lines are obtained with \texttt{MadGraph} at NLO, and the dashed lines are based on the tree-level result (\ref{eq:width}).  
A comparison of the measured top width \cite{ParticleDataGroup:2026mpi} with the SM prediction rules out the upper shaded region.
\vspace*{1mm}
} 
\label{fig:BR}
\end{figure}
%%%%%%

At the end of this Section we will estimate the current limit on the $\chi$ mass as $m_\chi \gtrsim 1.5$ TeV based on CMS and ATLAS searches for related vectorlike quarks. Those searches are mostly insensitive to the product of Yukawa couplings (although the ratio of these couplings determines the branching fractions of $\chi$), and are highly sensitive to $m_\chi$. The nonstandard top  branching fraction, on the other hand, scales as $\, y_{\chi \phi}^2 y_{\chi \hspace*{-0.3mm} { \scriptscriptstyle H} }^2/m_\chi^2$, underscoring the complementarity between the high-mass LHC searches, and the measurements of the top quark properties.

In Figure \ref{fig:BR}, we plot  the branching fraction for the $t \to \phi_t \, c$ decay as a function of $M_\phi$ for two sets of parameter values: 
$m_\chi = 1.5 $ TeV and $y_{\chi \phi} y_{\chi \hspace*{-0.3mm} { \scriptscriptstyle H} } = 1$ (blue lines), 
as well  $m_\chi = 2 $ TeV and $y_{\chi \phi} y_{\chi \hspace*{-0.3mm} { \scriptscriptstyle H} } = 0.3$ (red lines).
The first set approximately maximizes the $t \to \phi_t \, c$ width within this model, and the second set selects a vectorlike quark mass likely within the reach of the High-Luminosity LHC (HL-LHC) high-mass searches combined with a smaller product of Yukawa couplings likely within the reach of a Run 3 search for $t \to \phi_t \, c$.
This plot indicates that collider searches should target a sub-percent nonstandard branching fractions of the top quark.

The solid lines in Figure \ref{fig:BR} are obtained with our \texttt{MadGraph} simulation performed at NLO in QCD, and the 
dashed lines are based on the tree-level results (\ref{eq:width}) and (\ref{eq:SMtwidth}). 
The shaded region at the top of the plot (for a nonstandard top branching fraction above 25.8\%) is excluded at the 95\% CL by comparing the PDG \cite{ParticleDataGroup:2026mpi} average of top width measurements with the SM prediction \cite{Chen:2022wit} of 1.33 GeV. It should be mentioned though that the most precise measurement included in the PDG average was reported by CMS  \cite{CMS:2014mxl} based on some theoretical assumptions.
The more model-independent measurement of the top width performed by ATLAS \cite{ATLAS:2019onj} would allow a nonstandard top branching fraction as large as 54\%.

We assume that the interaction terms (\ref{eq:Yukawa}) are written in a field basis where the SM up-type quark matrix is already diagonal. 
The Higgs Yukawa term from (\ref{eq:Yukawa}) leads to an off-diagonal mass between the the vectorlike quark and SM up-type quark of the third generation, which we label $u^3$ to distinguish from the physical state $t$.
In order to account properly for this mass mixing, we label the gauge eigenstate of the vectorlike quark as $\chi^\circ$ (which is different from  the physical state $\chi$), and we write its mass term as $- m_{\chi^\circ} \overline \chi^\circ \chi^\circ$. With this more precise notation, the mass terms for  $u^3$ and $\chi^\circ$ take the following form:
\be
- \left(\overline u_{_L}^3  \, , \, \overline \chi_{_L}^\circ \right) 
\left(\!  \ba{cc} y_{_{33}} \, v_{\! _H}  \; & \;   y_{\chi \hspace*{-0.3mm} { \scriptscriptstyle H} } \, v_{\! _H}  \\ [2mm]   0 \; & \; m_{\chi^\circ}  \ea  \! \!  \right)
 \left(  \ba{c}  u_{_R}^3  \\ [2mm]   \chi_{_R}^\circ  \ea\!  \right)  + {\rm H.c.}  ~~,
\ee
where $y_{_{33}}$ is the Yukawa coupling of $u^3$ to the Higgs doublet. 
We diagonalize the above mass matrix following results from \cite{Chivukula:1998wd}. 

As typical LHC limits on vectorlike quark masses are in excess of 1 TeV, it is appropriate to expand in $v_{\! _H}^2 / m_{\chi^\circ}^2 \ll 1$.  
The top quark and $\chi$ masses are given by
\bear
&& m_t \approx  y_{_{33}} v_{\! _H}  \left( 1 -   \frac{ y_{\chi \hspace*{-0.3mm} { \scriptscriptstyle H} }^2 \, v_{\! _H}^2 }{2m_{\chi^\circ}^2} + O \left(v_{\! _H}^4 / m_{\chi^\circ}^4  \right)  \rule{0mm}{4mm} \right)  ~~,
\nonumber \\ [-1.5mm]
 \\ [-1.5mm]
&& m_{\chi} \approx  m_{\chi^\circ}  \left( 1 +  \frac{y_{\chi \hspace*{-0.3mm} { \scriptscriptstyle H} }^2  \, v_{\! _H}^2 }{2 m_{\chi^\circ}^2}
+ O \left( v_{\! _H}^4 / m_{\chi^\circ}^4  \right)  \rule{0mm}{4mm} \right)  ~~.
\nonumber
\eear
The relations between the gauge and mass eigenstates are
\bear
&& u^3_{_{L}} =  t_{_{L}} + \frac{y_{\chi \hspace*{-0.3mm} { \scriptscriptstyle H} } \, v_{\! _H} }{m_{\chi}}  \, \chi_{_{L}}   
+ O \left( v_{\! _H}^2 / m_{\chi}^2\right)   \;\;  ,
\nonumber \\ [-1.5mm]
\label{eq:relation} \\ [-1.5mm]
&& \chi_{_{L}}^\circ  = \chi_{_{L}} - \frac{y_{\chi \hspace*{-0.3mm} { \scriptscriptstyle H} } \, v_{\! _H} }{m_{\chi}} \,  t_{_{L}}    + O \left( v_{\! _H}^2 / m_{\chi}^2\right)    \;\; 
\nonumber
\eear 
for  left-handed fermions, and $u^3_{_{R}} = - t_{_{R}}$,  $\chi_{_{R}}^\circ  = \chi_{_{R}}$  for right-handed fermions, neglecting  corrections of order $v_{\! _H}^2 / m_{\chi}^2$.

Writing now more precisely the Lagrangian terms (\ref{eq:Yukawa}) in the gauge eigenstate basis 
($- y_{\chi \phi} \, \phi_t \,  \overline \chi_{_L}^\circ  c_{_R}  ...$), we use the relations between bases (\ref{eq:relation}) to find the interactions of the physical $\chi$ with the complex scalar $\phi_t$ and the SM Higgs boson $h^0$:
\be
- y_{\chi \phi} \, \phi_t \, \left( \overline \chi_{_L}   - \frac{y_{\chi \hspace*{-0.3mm} { \scriptscriptstyle H} } \, v_{\! _H} }{m_{\chi}} \, \overline  t_{_L}      \right) c_{_R}   
-  \frac{y_{\chi \hspace*{-0.3mm} { \scriptscriptstyle H} } }{\sqrt{2}} \, h^0 \, \left( \overline t_{_L}   +  \frac{y_{\chi \hspace*{-0.3mm} { \scriptscriptstyle H} } \, v_{\! _H} }{m_{\chi}} \, \overline  \chi_{_L}      \right)  \chi_{_R} 
+ {\rm H.c.} 
\label{eq:Yukawa-eigenstate}
\ee 
where we kept terms only up to  order  $v_{\! _H} / m_{\chi}$. Notice that the Yukawa coupling $\phi_t \overline t_{_L}   \chi_{_R}$ here is exactly the one from (\ref{eq:phitc}). The $\phi_t \overline \chi_{_L}   c_{_R}$ coupling induces the $ \chi \to \phi_t \, c$ decay with a 
tree-level width
\be
\Gamma( \chi \to \phi_t \,  c) \approx \frac{y_{\chi \phi}^2}{32 \pi} \, m_\chi
\label{eq:chinew}
\ee
up to small corrections of order $(m_t/m_\chi)^2$.

Similarly, the $h^0 \, \overline t_{_L}   \chi_{_R}$ coupling from (\ref{eq:Yukawa-eigenstate}) induces a $ \chi \to h^0 \, t$ decay. In addition, the SM couplings of $u^3$ to the $W$ and $Z$ bosons, written with (\ref{eq:relation}) in terms of the physical fermions, lead to the $ \chi \to W^+ \, b$ and $ \chi \to Z \, t$ decays, which have widths related to $ \chi \to h^0 \, t$ in the $m_\chi^2 \gg m_t^2$ limit \cite{Dobrescu:2009vz, Han:2003wu}:
\be
\Gamma ( \chi \to h^0 \, t) \approx  \Gamma ( \chi \to Z \, t) \approx  \frac{1}{2} \, \Gamma ( \chi \to W \, b) \approx  
\frac{y_{\chi \hspace*{-0.3mm} { \scriptscriptstyle H} }^2}{64 \pi} \, m_\chi  ~~.
\label{eq:chiSM}
\ee
Consequently, for  $y_{\chi \phi}^2 \gg y_{\chi \hspace*{-0.3mm} { \scriptscriptstyle H} }^2$ the dominant $ \chi$ decay mode is the nonstandard 
one $\chi \to \phi_t \,  c$ (see Section~\ref{sec:3body} for a discussion of the LHC phenomenology in this case). In the opposite limit, $y_{\chi \phi}^2 \ll y_{\chi \hspace*{-0.3mm} { \scriptscriptstyle H} }^2$, the standard modes  $\chi \to h^0 \, t$, $Z \, t$, $W\, b$ have the largest branching fractions.
In that case, the current mass limit set by CMS \cite{CMS:2022fck} is $m_\chi > 1.48$ TeV while the mass limit set by ATLAS is in the $1.36 - 1.4$ TeV range  \cite{ATLAS:2024gyc}; these limits are based on $\chi\bar \chi$ production, which is independent of the Yukawa couplings. When 
$y_{\chi \phi}$ and $ y_{\chi \hspace*{-0.3mm} { \scriptscriptstyle H} }$ have comparable values, the pair production signal is a combination 
of all above decays (a case of ``hybrid" signals discussed in \cite{Bernreuther:2025uwp}), and the mass limit is likely decreased because the backgrounds are larger (there are fewer leptons in the final state and also there fewer boosted $b\bar b$ pairs). 

Flavor-changing couplings involving the top quark can be usually constrained by measurements of rare meson processes  \cite{Fox:2007in} (see  \cite{Cremer:2023gne} for a different method). 
In our case, the effective coupling (\ref{eq:phitc}) does not lead to any tree-level flavor-changing processes other than the nonstandard top decay. At one loop there is one flavor-changing neutral current coupling that is not suppressed by off-diagonal CKM elements: $\phi_t \overline s \, b$ (see Section 3 for a detailed discussion of that loop-induced coupling). Given that we focus on the case of a complex scalar heavier than any meson, which furthermore has only one flavor off-diagonal coupling at tree level, $\phi_t$ does not contribute at one loop to any low-energy flavor process. 

The renormalizable theory that includes the vectorlike quark is also safe from low-energy flavor constraints. There is, however, the issue of why other gauge-invariant Lagrangian terms, such as $\phi_t \overline \chi_{_L} u_{_R}$ or  $\phi_t \overline \chi_{_L} t_{_R}$  are not included in (\ref{eq:Yukawa}). It turns out that such terms can be avoided by assigning charm number 1 to $\phi_t$.  
In that case  the  couplings (\ref{eq:Yukawa-eigenstate}) as well as the SM $H \,  \overline  q^2_{_L} c_{_R} $ term preserve charm number, while the absent terms mentioned above are excluded by a symmetry. 
Note that the SM mass matrix for the up-type quarks may be diagonal from the beginning, with all of the off-diagonal CKM entries originating from the down-type quark sector. 
Even so, terms of the type $ H^\dagger \overline  q^2_{_L} d^j_{_R} $ ($j$ labels the SM generations) would violate the charm symmetry, but their effects are suppressed by the small strange quark mass or by off-diagonal CKM entries. 

\bigskip

%%%%%%%%%%%%%%%%%%%%%%%%%%%%%%%%%%%%%%%%%%%%%%%%%%%%%%%%%%%%
%%%%%%%%%%%%%%%%%%%%%%%%%%%%%%%%%%%%%%%%%%%%%%%%%%%%%%%%%%%%
\section{Decays of the scalar and  vectorlike quark signals} 
\label{sec:3body}   \setcounter{equation}{0}

\begin{figure}[t!]
\vspace*{3mm}
\centerline{\hspace*{-.6cm}   \includegraphics[width=6.5cm, angle=0]{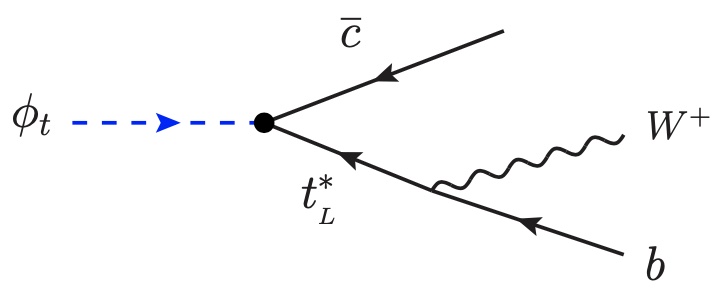}   }
\vspace*{-0.35cm}  
\caption{Three-body decay of the complex scalar $\phi_t$ through an off-shell left-handed top quark, due to 
interaction (\ref{eq:phitc}), which is depicted by the {\color{black} $\bullet$} symbol.
For this process to yield a $\overline c \, b \, W^+$ final state,  the mass of $\phi_t$  must satisfy the inequalities (\ref{eq:massrange}).
} 
\label{fig:diagramphi}
\end{figure}

In this Section we analyze the decay modes of the gauge-singlet complex scalar $\phi_t$. 
As discussed in Section~\ref{sec:model},  the nonstandard  2-body decay of the top quark, $t\to c \, \phi_t$, require $m_t  > M_\phi + m_c$. 
In addition, we consider the mass range of  $\phi_t$ that allows the 3-body decay of the complex scalar, induced by operator (\ref{eq:op5}),
via an off-shell top quark:
\be 
\phi_t \to t^* \overline c \to W^+ b \, \overline c   ~~.
\label{eq:process}
\ee
Note that the related process through an off-shell anti-top quark, $\phi_t \to \overline t^*  c \to W^+ \overline b \,  c $, is not induced by the new physics operator (\ref{eq:op5}). 
Combining the two mass conditions, the values of $M_\phi$ satisfy
\be
86 \; {\rm GeV} \approx  M_W + m_b + m_c < M_\phi <  m_t - m_c \approx 171 \; {\rm GeV}  ~~.\
\label{eq:massrange}
\ee
The Feynman diagram for the 3-body decay of $\phi_t$ is shown in Figure~\ref{fig:diagramphi}.

%%%%%%%%
%%%%%%%

 In the limit where the $b$ and $c$ quark masses are neglected, the width of the 3-body decay (\ref{eq:process}) 
 can be written as 
\be
\Gamma (\phi_t \to W b \, \overline c ) = \frac{1}{64\pi^3 M_\phi}  \, \int_0^{\mu_c}  d E_{\bar c}  \,   \int_{\mu_c - E_{\bar c}}^{E_b^{\rm max}} \! d E_b  \, 
\overline{\left| {\cal M} (\phi_t \!\to\! W b \, \overline c ) \right|^2}   ~~.
\label{eq:phase}
\ee
Here $E_{\bar c} $ and $E_{b}$ are the energies of the anti-charm and bottom quarks in the $\phi_t$ rest frame. The maximum value 
of $E_b$ for a given $E_{\bar c}$ is \cite{Dobrescu:2007xf}
\be
E_b^{\rm max} = \frac{1}{2} \left( M_\phi -  \frac{ M_W^2 }{M_\phi - 2 E_{\bar c}}  \right) ~~,
\ee
while the maximum value $E_{\bar c}$ for any  $E_b$ is
\be
\mu_c = \frac{1}{2} \left( M_\phi - \frac{M_W^2}{M_\phi}  \right)  ~~.
\ee
Ignoring the top-quark and $W$ boson  widths,  we obtain the following averaged squared matrix element for the  process (\ref{eq:process}):
\be \hspace*{-0.4cm}
\overline{\left| {\cal M} (\phi_t \!\to\! W b \, \overline c ) \right|^2}   =  3 \, \frac{ p_b \cdot  p_{\bar c}   \left[ 4 (p_b \cdot  p_{_W} \!)^2 -M_W^4 \right]  + 4 M_W^2  ( p_b \cdot  p_{_W}\!) \, p_{\bar c}  \cdot  (p_{_W}\! + p_b)  }{  \Lambda_5^2\,   \left( (p_{_W} \!+ p_b)^2 - m_t^2 \rule{0mm}{3.8mm} \right)^2}  ~~,
\ee
where again the $b$ and $c$ quark masses are neglected. Note that here we used the tree-level relation $M_W = g_w v_{\! _H}  /\sqrt{2}$.

The kinematics relations in the $\phi_t$ rest frame, {\it e.g.}, $p_b   \cdot  p_c = M_\phi (E_b + E_c - \mu_c)$, lead to the following form for the $\phi_t$ width from Eq.~(\ref{eq:phase}):
 \be
\Gamma (\phi_t \to W b \, \overline c ) = \frac{3 M_\phi^2}{32\pi^3 \Lambda_5^2}  \, \int_0^{\mu_c}  d E_{\bar c}  
\frac{f(E_{\bar c})}{ \left(m_t^2 - M_\phi^2 + 2  M_\phi  E_{\bar c}  \rule{0mm}{3.8mm} \right)^2} ~~,
\label{eq:phiGamma}
\ee
where the function of $E_{\bar c}$ is given by 
\be  \hspace*{-0.4cm} 
f(E_{\bar c}) = \int_{\mu_c \!-\! E_{\bar c}}^{E_b^{\rm max}} \!\! d E_b  \left[ \! \left( \! 4\left( E_{\bar c} - \mu_c\right)^2 - \frac{M_W^4}{M_\phi^2}  \right)      \left( E_b + E_{\bar c} - \mu_c\right)
- 4 \frac{M_W^2}{M_\phi}  E_{\bar c} \left( E_{\bar c}  \!-\!  \mu_c\right)  \rule{0mm}{3.mm}  \!  \right]  .
\ee
The integrals over $E_b$  and $E_{\bar c}$ can be performed analytically, and the final result for the width is
\bear  
&& \hspace*{-1.1cm} \Gamma (\phi_t \! \to \! W b \, \overline c ) = \frac{3 M_\phi^3}{2^8  \pi^3 \Lambda_5^2 \!} 
\left\{ \frac{r_\phi \! - \! r_{_W} }{r_\phi^3}  \! \left[2 +\! r_{_W}\! \! - \! \frac{7r_{_W}^2}{3}   \!  - \! \left(2 + \! \frac{5  r_{_W} \!}{6} \! - \! r_{_W}^2 \! \right) \! r_\phi \! + \! \frac{r_\phi^2}{6}   \rule{0mm}{4mm} \right] \! + \! \frac{ r_{_W}^3 }{ r_\phi \!} \ln \!\left(\!\frac{r_\phi }{ r_{_W}}\!\right) \right. 
\nonumber \\ [2mm]
&& \hspace*{3.3cm} \left. +  \frac{1}{ r_\phi\!} \left(\frac{2 \! - 3  r_{_W}^2 \! \! +  r_{_W}^3 \!}{r_\phi^2}  -  3\frac{1 \! - r_{_W}^2 \!}{r_\phi} \!  + \! 1\! -  \!  r_{_W}^3 \! \! \right)  \ln \! \left( \frac{1 \!-\! r_\phi \!}{1 \!-\! r_{_W} \!} \right) \! \right\}
~,
\label{eq:finalWidth}
\eear
where we defined 
\be
r_\phi = \frac{M_\phi^2}{m_t^2} \;\;\:  , \;\;\;
r_{_W} \! = \frac{M_W^2}{m_t^2} \approx 0.217     ~~.
\ee 

For $\Lambda_5 = 1.5$ TeV, we obtain $\Gamma (\phi_t \to W b \, \overline c ) \approx  8.6$ eV for $M_\phi = 100$ GeV, and 7.9 keV  for $M_\phi = 150$ GeV.
The 3 orders of magnitude increase in the width when $M_\phi$ varies in that range is due to the large phase-space suppression when $M_\phi $ approaches $M_W$. 
In that limit, the  analytic form of the width (\ref{eq:finalWidth}) allows us to derive the dependence on $M_\phi$  as  
\be
\Gamma (\phi_t \! \to \! W b \, \overline c ) = \frac{3 M_W^2}{160 \pi^3 \Lambda_5^2 \left(m_t^2 - M_W^2\right)^2} \;
(M_\phi - M_W)^5  \;\; , \;\; {\rm for} \;  M_\phi - M_W \ll M_W~~.
\ee

%%%%%%%%
\begin{figure}[t!]
\centerline{\hspace*{-.6cm}   \includegraphics[width=10.5cm, angle=0]{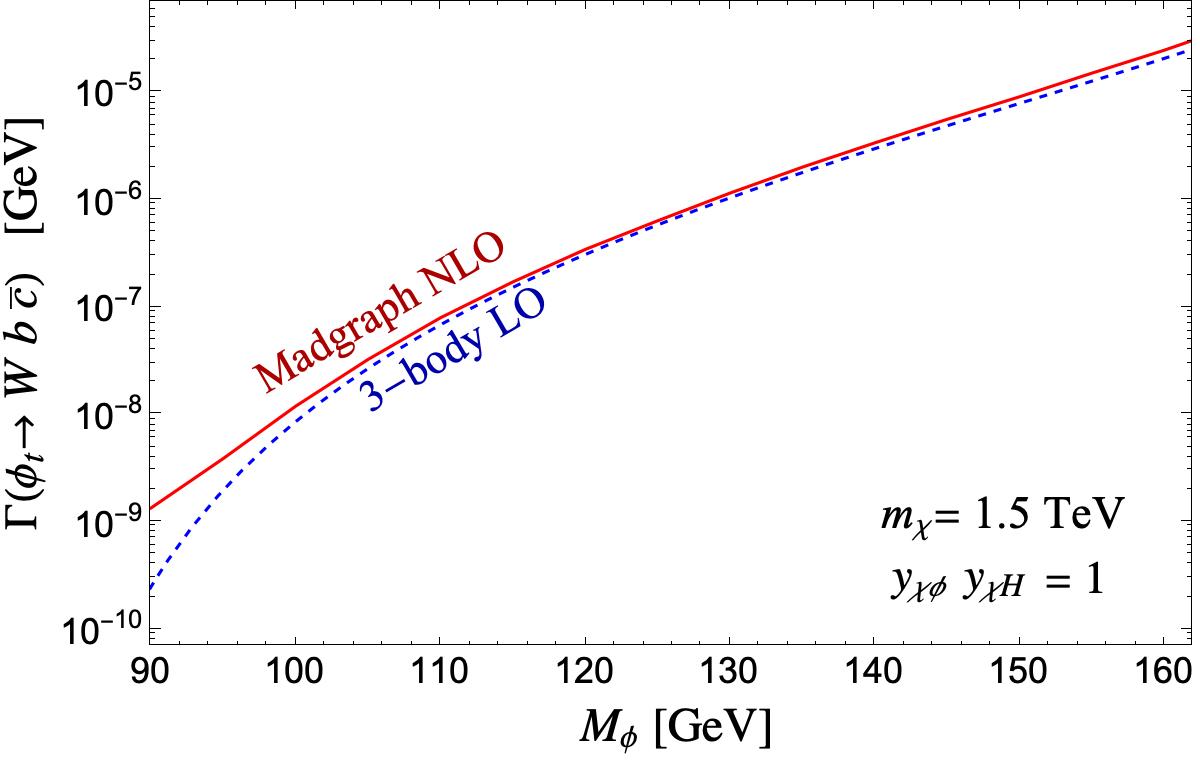}}   
\vspace*{-0.1cm}  
\caption{Width of the complex scalar $\phi_t$ as a function of the $\phi_t$ mass, due to the process $\phi_t \to W^{(*)} b \, \overline c$  that proceeds through an off-shell top quark. Solid red line was obtained with the \texttt{MadGraph} simulation at NLO in $\alpha_s$ with $b$ and $c$ masses taken into account, and allowing an off-shell $W$ boson ({\it i.e.}, a 4-body $\phi_t$ decay).  Dashed blue  line is based on the 3-body tree-level analytic result (\ref{eq:finalWidth}). 
The dimension-5 operator (\ref{eq:op5}), responsible for this decay, is induced by a vectorlike quark  of mass  $m_\chi = 1.5$ TeV with a Yukawa coupling product  $y_{\chi \phi} y_{\chi \hspace*{-0.3mm} { \scriptscriptstyle H} } = 1$. 
\vspace*{4mm}
} 
\label{fig:width}
\end{figure}
%%%%%%%

We emphasize that the $b$ and $c$ quark masses have been neglected in the above derivation, and thus the correct 3-body width is substantially smaller than (\ref{eq:finalWidth})  for $M_\phi$ below  $\sim 95$ GeV. 
On the other hand, as $M_\phi$ decreases towards $M_W + m_b + m_c$, it becomes more likely that $\phi_t $ decays through an off-shell top and then through an off-shell $W$. 
These 4-body decays, $\phi_t \to b  \, \overline c \, c \overline s $, $\phi_t \to b  \, \overline c \, u \overline d $ and $\phi_t \to b  \, \overline c \, \ell^+ \nu $, increase the $\phi_t$ width compared to the 3-body result.

In Figure~\ref{fig:width} we show $\Gamma (\phi_t \to W^+ b \, \overline c )$ as a function of the $\phi_t$ mass, for a product of Yukawa couplings  $y_{\chi \phi} y_{\chi \hspace*{-0.3mm} { \scriptscriptstyle H} } = 1$, and  a vectorlike quark mass $m_\chi = $ 1.5 TeV.
 The dashed  line represents the  analytic result (\ref{eq:finalWidth}) with $b$ and $c$ masses neglected, 
 while the solid line (which is clearly higher at $M_\phi \lesssim 100$ GeV) represents a \texttt{MadGraph} \cite{Alwall:2014hca} simulation performed at  NLO in the QCD coupling constant ($\alpha_s$) with default quark masses ($m_b = 4.7$ GeV and $m_c = 1.3$ GeV). 
 The \texttt{MadGraph} model files  with the $\phi_t$ complex scalar and its effective Yukawa interaction (\ref{eq:phitc}) were generated by FeynRules \cite{Alloul:2013bka} as described in Section~\ref{sec:model}. 
 The multi-body decay rate is computed by the MadWidth \cite{Alwall:2014bza} module of  \texttt{MadGraph}.
 
\begin{figure}[t!]
\centerline{\hspace*{-.6cm}   \includegraphics[width=6.5cm, angle=0]{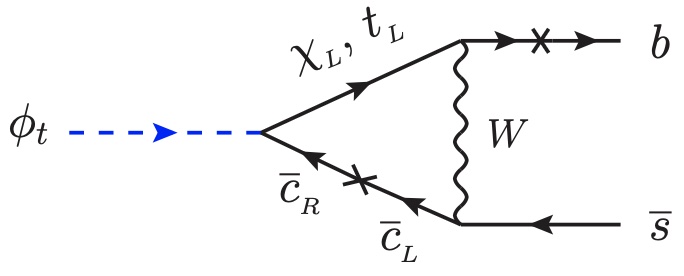}   }
\vspace*{-0.4cm}  
\caption{One-loop diagram responsible for the main 2-body decay of the complex scalar: $\phi_t \to b \bar s$.  The two $\times$ symbols represent mass insertions for the charm and bottom quark lines. The Yukawa interactions of $\phi_t$ with $\overline \chi_{_L} c_{_R}  $ and $  \overline  t_{_L}  c_{_R}$  are given in (\ref{eq:Yukawa-eigenstate}).
\vspace*{1mm}
} 
\label{fig:loop}
\end{figure}

An alternative decay mode is the 1-loop induced process $\phi_t \to b \, \overline s$, which occurs through two diagrams with a $W$ boson, a charm quark, and either the $t$ or $\chi$ mass eigenstates running in the loop (see Figure~\ref{fig:loop}).
The Yukawa interactions that appear here are given by the first two terms in (\ref{eq:Yukawa-eigenstate}).
Note that the $c$ quark in the loop requires a chirality flip, and the external $b$ quark line also involves a mass insertion. Thus, the 1-loop width is suppressed by $m_c^2 \, m_b^2$, where the quark masses are evaluated at the $M_\phi$ scale.  
The logarithmic divergences of the $t$ and $\chi$ loops cancel each other, so the result is finite, as expected in a renormalizable theory.  

Parametrically, the width for the 1-loop decay is 
\be
\Gamma(\phi_t \to b \, \overline s) = \frac{3 g_w^2 \, m_c^2 \, m_b^2 }{2^{13} \pi^5 \Lambda_5^2 \, M_W^2}  
\, M_\phi \; {\cal F}(M_\phi, m_\chi)    ~~.
\ee
Here we included a typical  loop factor squared $(4\pi)^{-4}$, the 2-body phase space $(4\pi)^{-2}$, the $W$ coupling squared $(g_w/\sqrt{2})^2$ (another squared coupling is removed by a $M_W^2/v_{_H}^2$ factor), as well as a dimensionless function $ {\cal F}$ that depends on 
$M_\phi$ and $m_\chi$. In order to  compute $ {\cal F}$, we used the FeynCalc package  \cite{Shtabovenko:2016sxi}.
In the $m_t^2 \ll m_\chi^2$ limit, we find 
 \be
 {\cal F}(M_\phi, m_\chi) = \frac{1}{4} \left( \ln\left(\frac{m_\chi^2}{m_t^2}\right) + O(1) + O\left(M_\phi^2/m_t^2\right)\rule{0mm}{3.9mm} \right)^{\! 2}  ~~.
 \ee
As expected for the rate of a 1-loop process, the leading logarithmic term in ${\cal F}$ is of order $\ln^{2\!}(m_\chi^2/m_t^2) > 19$.
For $M_\phi \lesssim 100$  GeV and   $\Lambda_5 \approx m_\chi \approx 1.5$ TeV, the width of the 1-loop process is  
\be
\Gamma(\phi_t \to b \, \overline s) \approx  5 \times 10^{-5} \; {\rm eV}  \left(\frac{M_\phi}{100  \, {\rm GeV}} \right) ~~,
\ee
which is 4 orders of magnitude smaller than the 3-body width at $M_\phi = 90$ GeV (see also Figure~\ref{fig:width}). 
Hence, only the tree-level decay through an off-shell top quark is relevant for the mass range (\ref{eq:massrange}) studied here.

%%%%%%%%
%%%%%%%

Since the total width of $\phi_t$ is $\Gamma_\phi \approx \Gamma (\phi_t \! \to \! W b \, \overline c )$, 
the value of the width obtained with \texttt{MadGraph} for  $M_\phi = 90$ GeV and $\Lambda_5 = 1.5$ TeV, namely $\Gamma_\phi \approx 1.4$ eV, implies that the decay length of $\phi_t$ in its center-of-mass frame is $c\tau_\phi \approx 1.4 \times 10^{-4}$ mm. As the decay length decreases fast at larger $M_\phi$, and the boost of $\phi_t$ at the LHC is at most of order 10, we conclude that the $\phi_t$ decays are always prompt  for $\Lambda_5 \lesssim 50$ TeV. 

%%%%%%%%
\begin{figure}[t!]
%\vspace*{3mm}
\centerline{\hspace*{-.6cm}   \includegraphics[width=9.6cm, angle=0]{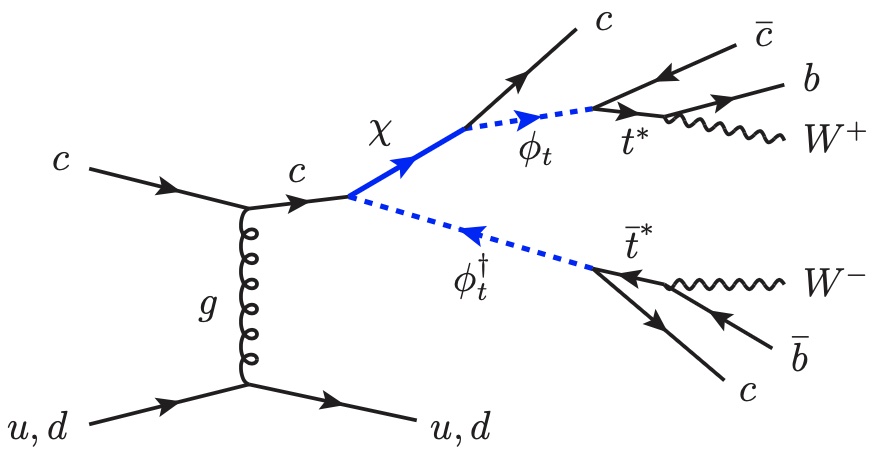}   }
\vspace*{-0.1cm}  
\caption{Single production of the vectorlike quark $\chi$ in association with a $\phi^\dagger_t$ scalar at the LHC, followed by cascade decays due to the first terms of interactions (\ref{eq:Yukawa-eigenstate}): $\chi \to \phi \, c$,  as well as $\phi_t$ and $\phi^\dagger_t$ decays through an off-shell $t$ and $\bar  t$, respectively.  An additional diagram with a gluon instead of a $u$ or $d$ quark is also important.
} 
\label{fig:chiphi}
\end{figure}
%%%%%%%

At the LHC, the vectorlike quark can be produced in pairs through its QCD coupling. 
The NLO cross section for that process at $\sqrt{s} = 13.6$ TeV decreases from 2 fb at $m_\chi = 1.5$ TeV to 0.4 fb at $m_\chi = 1.8$ TeV \cite{Dobrescu:2024mdl}, and is independent of any other parameters. 
Single $\chi$ production in association with a $\phi_t$ is also possible due to the first interaction in (\ref{eq:Yukawa}).
That process has a sizable cross section, but it is more model dependent: 
it scales as $y_{\chi \phi}^2$ and it depends on both $m_\chi$ and $M_\phi$. 
By far the largest contributions  come from an initial state with a charm quark plus either a $u$ or $d$ quark (shown in Figure~\ref{fig:chiphi}) or a gluon (same topology as in Figure~\ref{fig:chiphi}, with the $u/d$ lines replaced by gluons). Besides this $\chi \phi_t^\dagger j$  production, where $j$ is a gluon or  light quark jet, there is $\overline \chi \phi_t j$  production, with equal cross section. Thus, there are two dominant cascade decays:
\bear
&& p p \to \chi \phi_t^\dagger j \to  (c \, \phi_t) \phi_t^\dagger j \to (c \overline c \, b W^+)(c \overline b W^-) j ~~,
\nonumber \\ [-1.5mm]
 \\ [-1.5mm]
 \nonumber 
&& p p \to \overline \chi \phi_t j \to  (\overline c \, \phi_t^\dagger) \phi_t j \to ( \overline c c \, \overline b W^-)(\overline c b W^+) j ~~,
\eear
which collectively we refer to as $\chi \phi_t \, j \to  3c + b \, \bar b \, W W j$  processes.

For a benchmark point in the parameter space, $y_{\chi \phi} = 0.6$ and  $M_\phi = 100$ GeV, the sum of the 
$\chi \phi_t \, j$ cross sections decreases  from approximately 75 fb at $m_\chi = 1.5$ TeV to 26 fb at $m_\chi = 1.8$ TeV. 
These are large enough rates for searches at the  ATLAS and CMS experiments. The $cbW$ system originating from the $\phi_t$ produced in a $\chi$ decay is highly boosted; in the case of hadronic $W$ decays, this yields a narrow jet (labelled here $J_{cbW}$) with 4-prong substructure and mass near $M_\phi$. 
Events in which the other $W$ decays leptonically appear as an $\ell cc b j \, J_{cbW}  + \met$ signal, which offers various opportunities for reducing the backgrounds.

As there are twice as many degrees of freedom in the Higgs doublet compared to $\phi_t$, from (\ref{eq:chinew}) and  (\ref{eq:chiSM}) follows that the branching fraction of  $\chi \to c \phi_t$ is 
\be
{\cal B} (\chi \to c \phi_t) \approx  \frac{y_{\chi \phi}^2} {y_{\chi \phi}^2 + 2 y_{\chi \hspace*{-0.3mm} { \scriptscriptstyle H} }^2}  ~.
\ee
Consequently, for a ratio of Yukawa couplings $y_{\chi \phi}/ y_{\chi \hspace*{-0.3mm} { \scriptscriptstyle H} }  > 2$, the branching fraction for the cascade process 
$\chi \phi_t \, j \to  3c + b \, \bar b \, W W j$ is above 67\%. 
In the remaining $\chi \phi_t \, j$ events the vectorlike quark decays into SM particles $\chi \to h^0 \, t$, $Z \, t$, $W\, b$, leading to several interesting final states. The fraction of events involving a $h^0 \, t$, $Z \, t$, or $W\, b$ system plus $\phi_t j$ increases for smaller $y_{\chi \phi}/ y_{\chi \hspace*{-0.3mm} { \scriptscriptstyle H} }$, and may also be included in searches at the LHC.

\bigskip

%%%%%%%%%%%%%%%%%%%%%%%%%%%%%%%%%%%%%%%%%%%%%%%%%%%%%%%%%%%%
%%%%%%%%%%%%%%%%%%%%%%%%%%%%%%%%%%%%%%%%%%%%%%%%%%%%%%%%%%%%
\section{Nonstandard top quark decays at the LHC}
\label{sec:LHC}  \setcounter{equation}{0}

\begin{figure}[b!]
\centerline{\hspace*{0.6cm}   \includegraphics[width=9.6cm, angle=0]{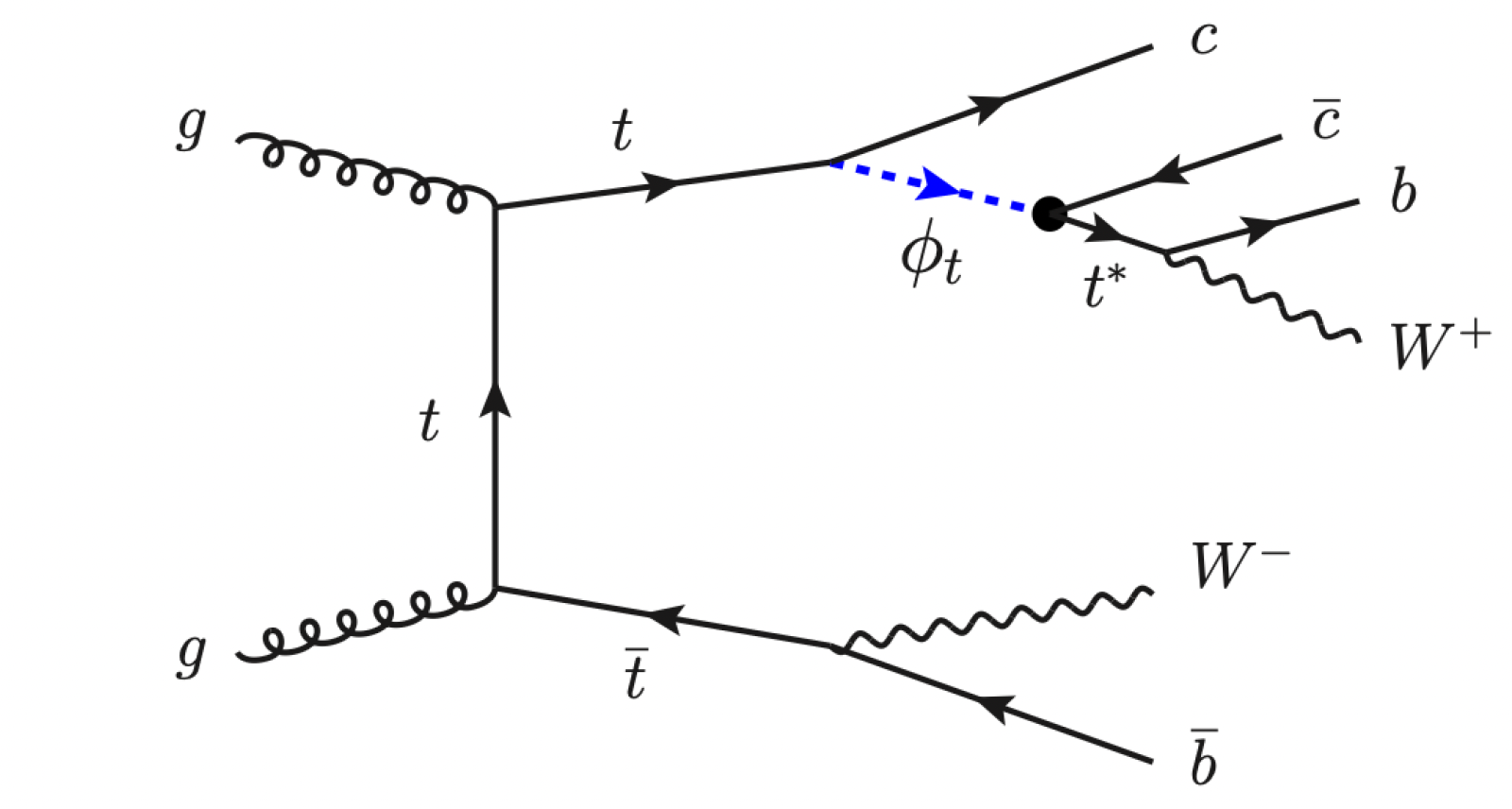}    }
\vspace*{-0.1cm}  
\caption{LHC process with a nonstandard top-quark decay, $t \to c\,\phi_t$, followed by the scalar decay 
$\phi_t \to \bar c\, t^{*} \to \bar c\, b W^+ $ through an off-shell top quark.
The $\bar t$ decays here as in the SM, so the final state is $(W^+ b\,\bar c\, c)(W^- \bar b)$.
The  {\color{black} $\bullet$} vertex represents the dimension-5 interaction (\ref{eq:phitc}).
A similar diagram with a nonstandard $\bar t$ decay leads to the $(W^+ b)(W^- \bar b\, c\, \bar c)$ final state.
\vspace*{1mm}}
\label{fig:pair}
\end{figure}

The top quark is mainly produced in pairs at the LHC through its QCD coupling. The main nonstandard processes for the mass of $\phi_t$ in the range $86~{\rm GeV}<M_{\phi}<171~{\rm GeV}$ are
\bear
&& p p \to t \overline t \to ( \phi_t \,  c) (W^- \overline b) \to ( W^+ b  \, \overline c \, c) (W^- \overline b)  ~~, % \;\; {\rm or} \;\; 
\nonumber \\ [-2mm]
\\ [-2mm]
\nonumber && p p \to t \overline t \to (W^+ b) ( \phi_t^\dagger \, \overline c)  \to ( W^+ b ) (W^- \overline b \, c \, \overline c)  ~~,
\eear
where the nonstandard decay has a branching fraction at the percent level, as derived in Section~\ref{sec:model}. The first of these two processes is shown by the Feynman diagram in Figure~\ref{fig:pair}.

The $t\bar t$ production total cross section in $pp$ collisions at $\sqrt{s} = 13.6$ TeV is $\sigma_{t\bar t} \approx 9 \times 10^5~{\rm fb}$ with both a theoretical and experimental uncertainty of approximately 4\%,  based on SM predictions~\cite{Czakon:2011xx} and measurements by the ATLAS~\cite{ATLAS:2023slx} and CMS~\cite{CMS:2023qyl} collaborations, respectively. Consequently, during Run 3 of the LHC, which accumulated ${\cal L}_3 = 330~{\rm fb}^{-1}$ of integrated luminosity, the number of $\phi_t$ or $\phi_t^\dagger$ produced in each of the two detectors was 
\be
N_{\phi_t} = 2 \sigma_{t\bar t} \,  {\cal L}_3 \,  {\cal B}(t \to \phi_t \, c) \approx  6 \times 10^8 \, {\cal B}(t \to \phi_t \, c)   ~~.
\ee
For $M_\phi = 100$ GeV and $\Lambda_5 = 1$ TeV, $y_{\chi \phi} = y_{\chi H}=1$, the nonstandard branching fraction is ${\cal B}(t \to \phi_t \, c) \approx 1.5\%$ (see Figure~\ref{fig:BR}), so that in each of the detectors the expected number of top pairs with a nonstandard top decay is ${\cal O}(10^7)$. At the high-luminosity LHC, this number is expected to grow by about an order of magnitude.

While the expected number of signal events is quite large, a search for the exotic $t\rightarrow \phi_t c$ decay must contend with the significant QCD background. 
The SM top pair production and decay results in a final state consisting of $W^+W^-b\bar{b}$, while the nonstandard processes result in the same particles but with two additional charm quark jets.
This can be faked in the SM if there is additional radiation in the standard $t\bar{t}$ production process that leads to two additional jets.
The cross section for such a background, $t\bar{t}+2j$ where $j$ is an inclusive jet of a gluon or any quark flavor, has been measured to be $292~{\rm pb}$~\cite{CMS:2020grm} at $\sqrt{s}=13~{\rm TeV}$.
However, if the additional jets can be tagged as charm quarks, then the background is significantly lower: CMS~\cite{CMS:2020utv} 
has determined the total cross section $\sigma_{t\bar{t}+c\bar{c}}$ to be approximately  $10.1~{\rm pb}$ at $\sqrt{s}=13~{\rm TeV}$.
In principle, this measurement (or a similar result from ATLAS~\cite{ATLAS:2024plw}) can constrain the branching fraction of our exotic decay mode; setting such a limit on our model requires determining the kinematic distributions  of the jets and leptons for the measurement's fiducial region.

To explore the observable signatures of how $\phi_t$ might manifest from top decays at the LHC, we have written our model as a UFO~\cite{Degrande:2011ua} model file using FeynRules~\cite{Alloul:2013bka}. We simulate $pp$ collisions at the LHC using \texttt{MadGraph}~\cite{Alwall:2014hca} at leading order, with the default PDF \cite{Ball:2013hta}, and discuss some distributions relevant for an LHC search as follows.

The transverse momentum ($p_T$) distributions of the charm and bottom quarks provide an important handle for separating signal from background, as they must be tagged to reduce the SM background.
Heavy flavor tagging for charm and bottom quarks has been explored by ATLAS~\cite{ATLAS:2024plw,ATLAS:2018bpl,ATLAS:2019bwq} and CMS~\cite{CMS:2020utv,CMS:2021scf}, and a minimum transverse momentum of the bottom and charm jets, $p_T^{(b,c)}\gtrsim20$ GeV, is typically required. 
Due to the mass of $\phi_t$ being somewhat comparable to the $W$ and top masses, the heavy flavor jets will be produced with relatively small momenta that, for a sizable portion of phase space, will not pass the minimum $p_T$ requirements typically employed for flavor tagging.

\begin{figure}
\centering
\includegraphics[width=0.5\linewidth]{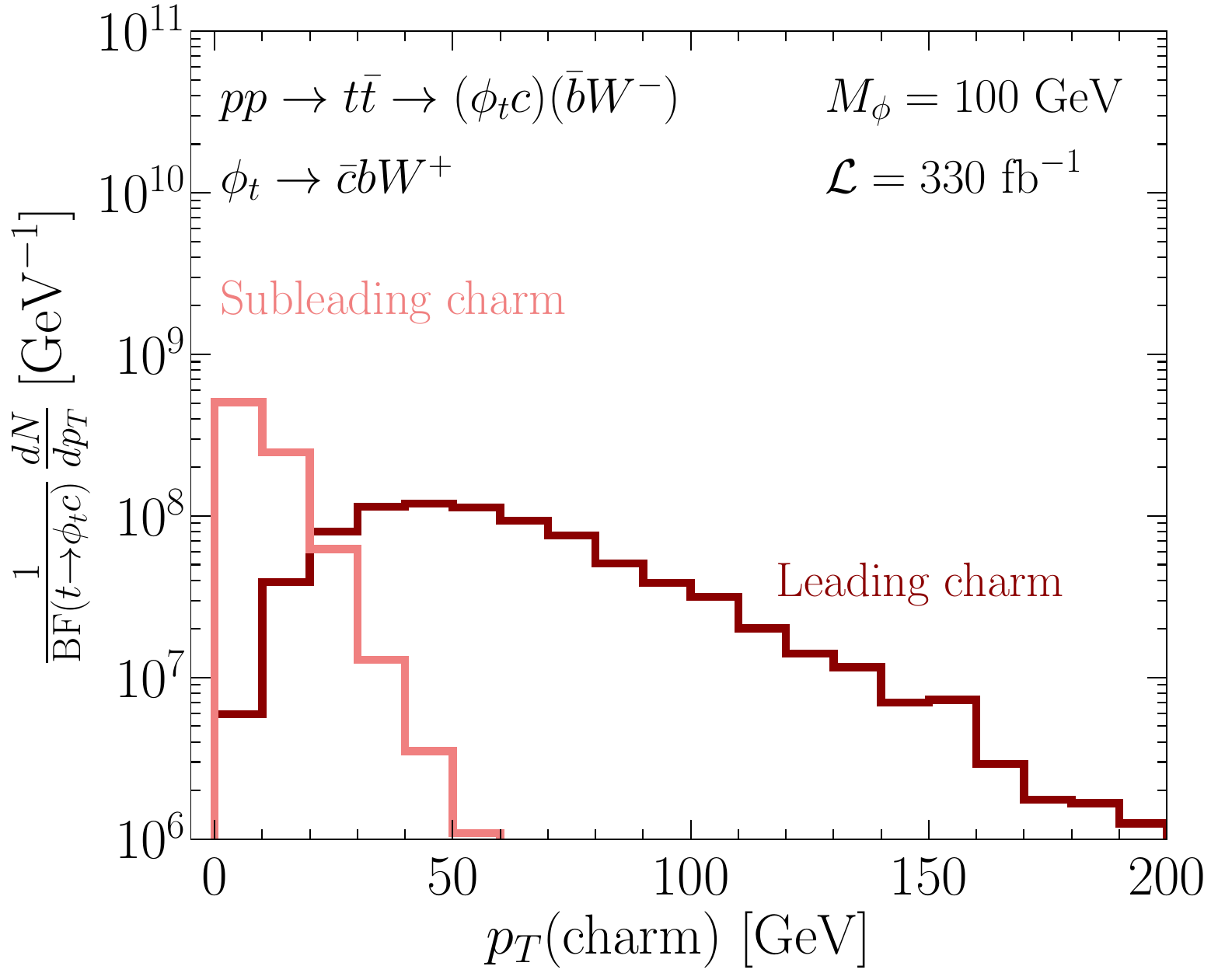}
\includegraphics[width=0.475\linewidth]{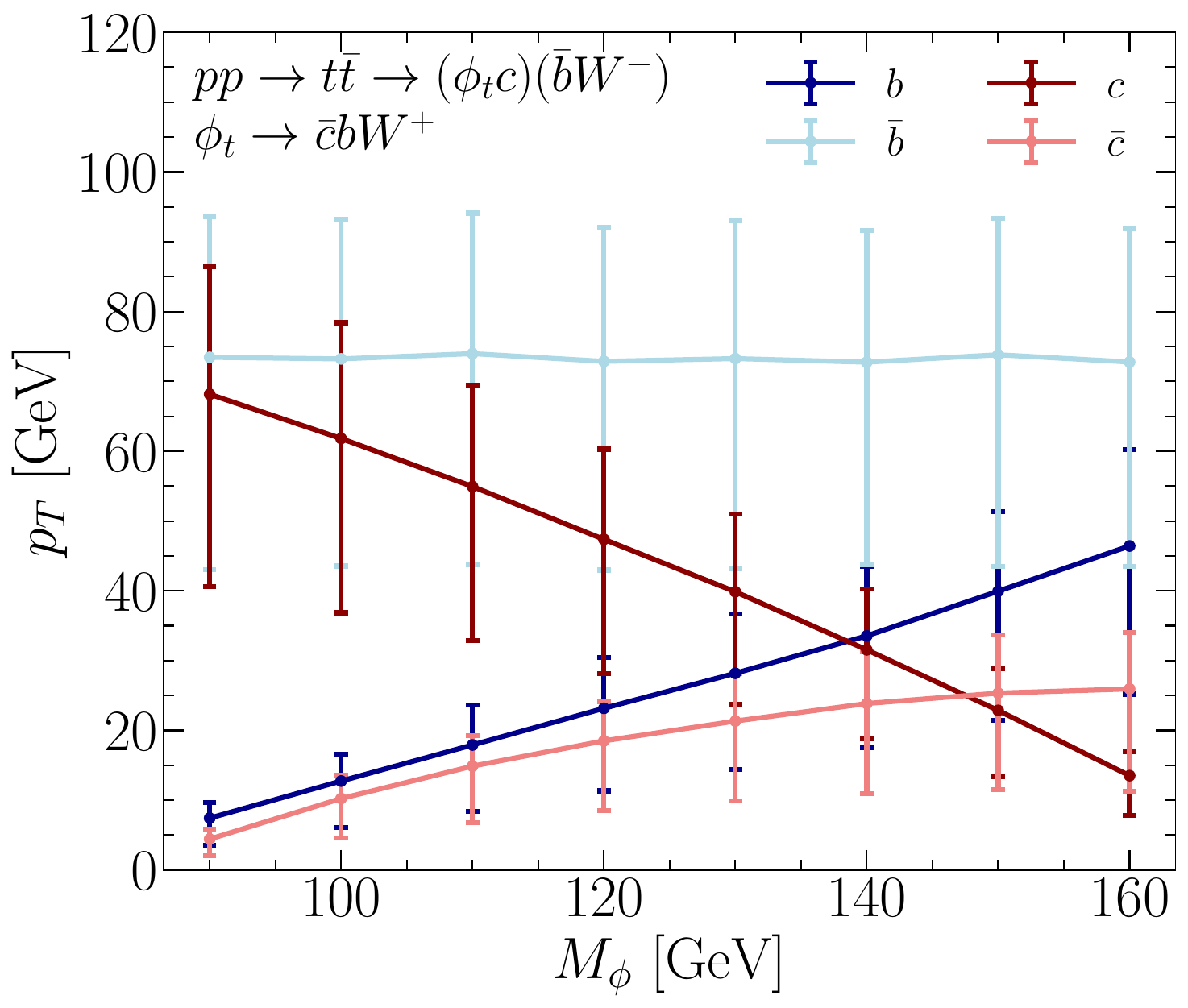}
\caption{\textit{Left:} Transverse momentum distributions of the leading (dark red) and subleading charm  (light red) quarks from $t\overline t$  production, where the top quark decays to $c \, \phi_t$ as in Figure~\ref{fig:pair} with $M_{\phi} = 100~{\rm GeV}$. \textit{Right:} For the same process, the spread in transverse momenta for each of the four heavy quarks as a function of $M_{\phi}$. The central data point is the average $p_T$, and the error bars denote the 25th and 75th percentiles. The dark blue (red) curve is for the bottom (charm) quark, while the light blue (red) is for the anti-bottom (anti-charm) quark.}
\label{fig:Charmpt}
\end{figure}

In the left panel of Figure~\ref{fig:Charmpt} we show the $p_T$ distribution of the two charm quarks for the case where the top quark decays through the nonstandard mode, \textit{i.e.}, $pp\rightarrow t\bar{t}\rightarrow (\phi_t c)(\bar{b}W^-)$, with $M_{\phi}= 100~{\rm GeV}$ in $pp$ collisions at $\sqrt{s}=13.6~{\rm TeV}$.
Typically, the leading charm (defined on a per-event basis as the charm quark with larger $p_T$ of the two) is the one associated with the decay of the top quark and the subleading charm quark is from the decay of $\phi_t$. 
For $M_{\phi}=100~{\rm GeV}$, the leading charm has an average $p_T$ of about 60~GeV and lies well within the fiducial phase space of the CMS $t\bar{t}+c\bar{c}$ measurement~\cite{CMS:2020utv}, which required $p_T > 20~{\rm GeV}$. The subleading charm is much softer, with an average $p_T$ of about 10~GeV, and exceeds 2~GeV in only about 9\% of events. The bottom-quark distributions are similar, except that the leading bottom quark has a higher $p_T$ than the leading charm quark because $M_W<M_{\phi}$.
%'

The resulting kinematics of the outgoing jets can vary substantially as $M_{\phi}$ is increased towards the top mass threshold.
In the right panel of Figure~\ref{fig:Charmpt} we show the average $p_T$ of the charm and bottom quarks, and their 25th and 75th percentiles as error bars, for $90~{\rm GeV} \leq M_{\phi} \leq 160~{\rm GeV}$.
As the mass of $\phi_t$ is increased, the leading $b$ quark $p_T$ does not vary significantly, since it comes from the SM decay process, while, on the other hand, the quarks that result from $\phi_t$ decay generally become more boosted. 
For larger $M_{\phi}$, the charm quark that was produced directly from the top decay becomes softer, and eventually becomes the subleading charm jet as the top quark mass threshold is approached around $M_{\phi}\sim 150~ {\rm GeV}$ (this turnover can be estimated by approximately equating the energies of the two charm quarks, $m_t-M_{\phi}=(M_{\phi}-M_W)/3$).
In our simulation, we find that approximately $2.3\%$ ($0.02\%$) of signal events have all four quark jets satisfying $p_T > 20$ ($50$)~GeV for $M_\phi = 100~{\rm GeV}$.
%%%%%%%%%%%%%%%%%%%%%%%%%%%%%%%%%%%%%%%%%%%%%%%%%%%%%%%%%%%%

\begin{figure}[t]
\centering
\includegraphics[width=0.49\linewidth]{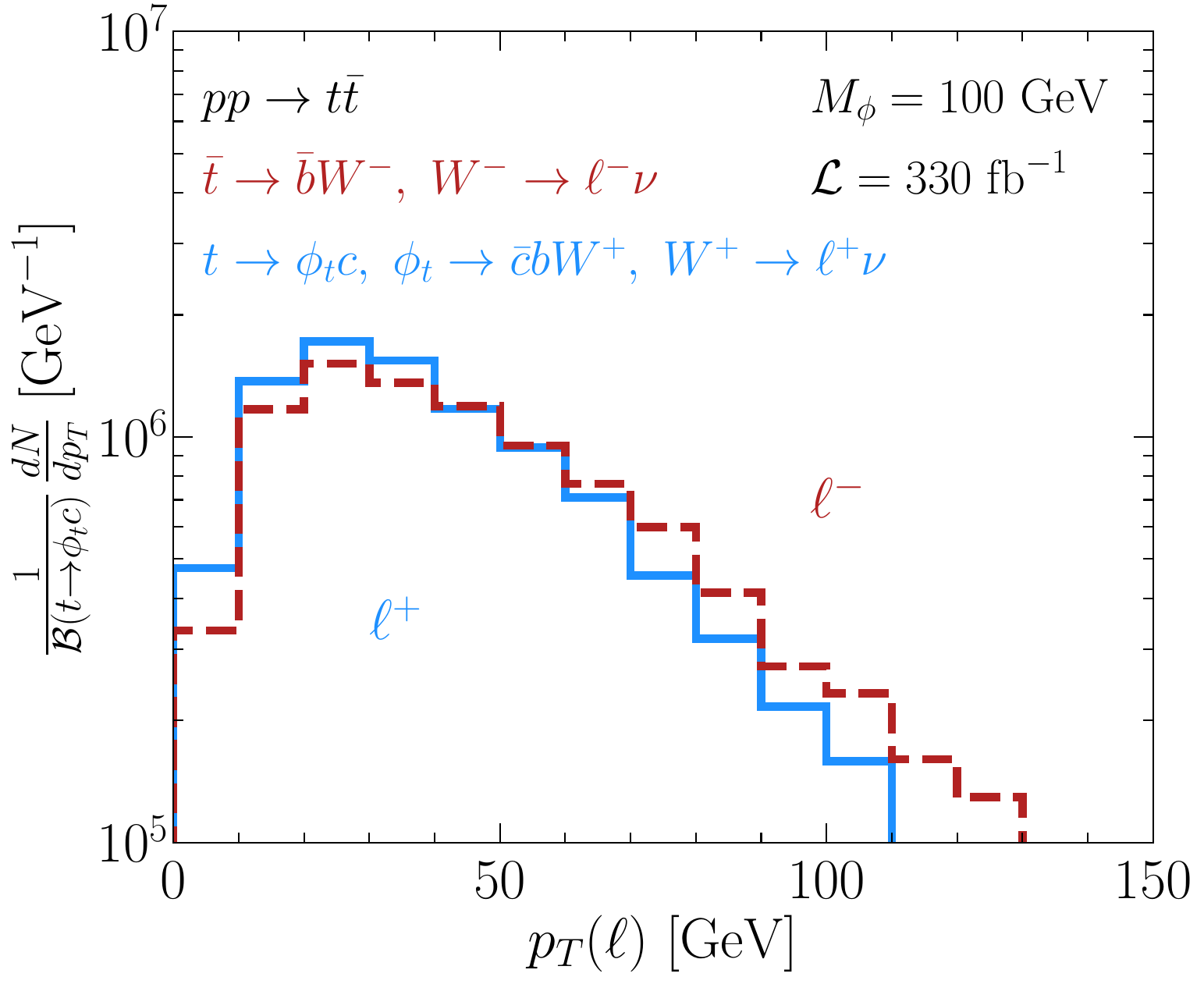}
\includegraphics[width = 0.50\linewidth]{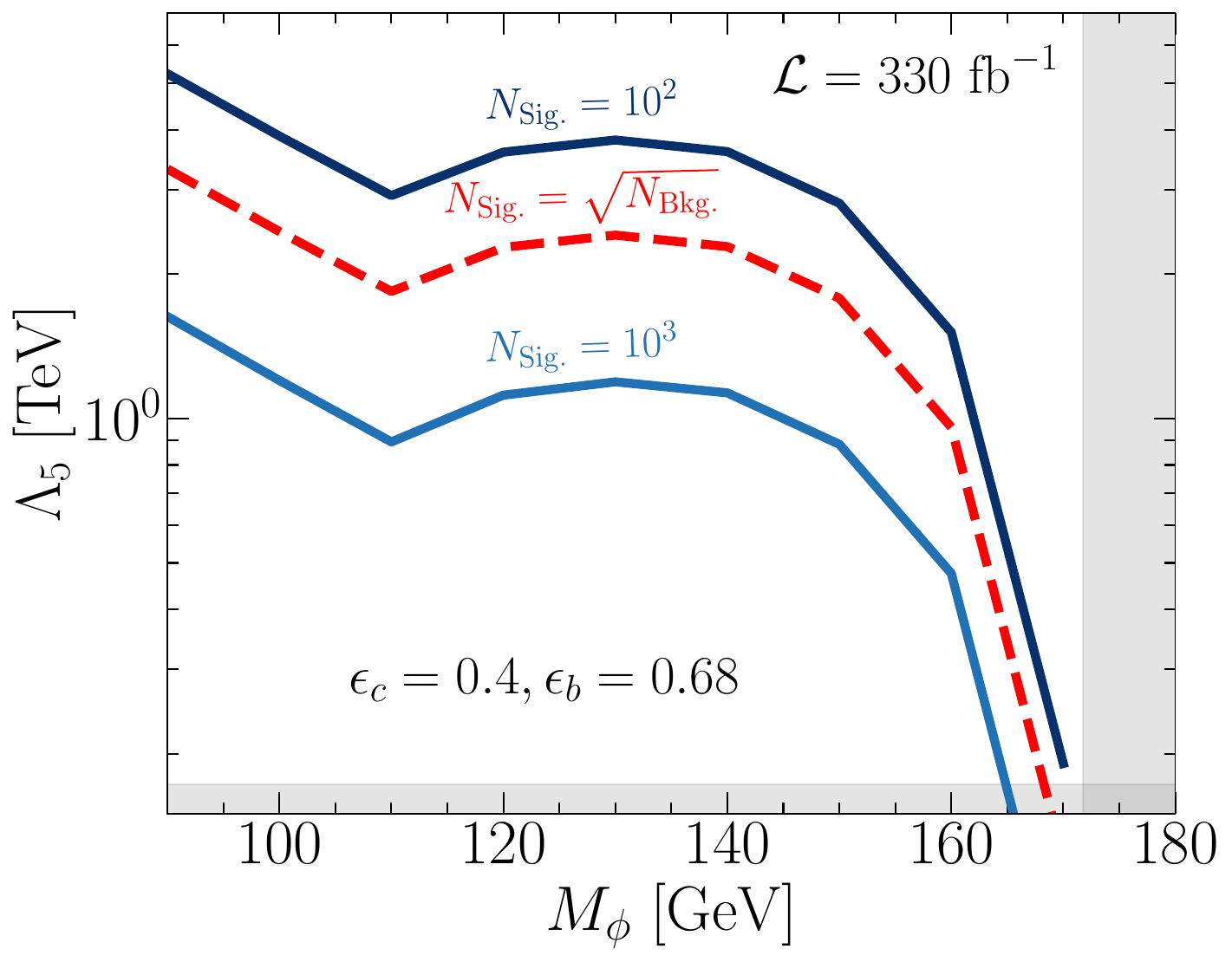}
\caption{\textit{Left:}  Parton-level $p_{\rm T}$ distribution, simulated in \texttt{MadGraph}, of the two leptons resulting from $t \bar t$ production, where the top quark decays to $c \phi_t$ for $M_{\phi}=100~{\rm GeV}$. The negatively charged lepton (dashed red) is from the standard anti-top decay and the positively charged lepton (solid blue) is from the nonstandard decay, and is softer due to the small $M_{\phi} - M_W$. 
\textit{Right:} Contours of constant event rate in the $\Lambda_5$ vs. $M_\phi$ plane, with cuts on lepton and jet kinematics (see text for details), for charm and bottom tagging efficiencies of $\epsilon_c=0.4,~\epsilon_b=0.68$. The dashed red curve is where the signal event rate equals the square root of the expected background, and serves as a sensitivity estimate for a background-only hypothesis. The gray-shaded regions label $M_\phi<M_W$ (the $W$ from the $\phi_t$ decay is off-shell) or $\Lambda_5<m_t$ (effective interaction is nonperturbative).}
\label{fig:Nsig}
\end{figure}

In addition to the jets, the kinematics of the outgoing leptons play an important role in identifying the signal, and a direct search may benefit from the differing kinematics of the two leptons.
When the top quark decays through the nonstandard channel involving $\phi_t$, the reduced phase space leaves the $W$ from the $\phi_t$ decay with lower momentum than in the standard $t \to bW$ decay, so the lepton from its decay is also softer.
Since the leptonic $W^\pm$ decay channel is likely to be the most constraining, we show the $p_T$ distributions of the outgoing leptons in the left panel of Figure~\ref{fig:Nsig} for $M_{\phi}=100~{\rm GeV}$.
Here the top decays to $\phi_t$, so the positively charged lepton originates from the nonstandard decay; its distribution, shown in solid blue, is shifted to lower transverse momenta.
The dashed red curve shows the $p_T$ distribution of the negatively charged lepton, which comes from the standard $t \to bW$ decay and thus follows the same distribution as the lepton in the fully SM process.

With the $p_T$ distributions in hand, we can now estimate the number of signal
events that pass a realistic set of selection cuts.
As a benchmark, we follow Ref.~\cite{CMS:2020utv} and require that:
\textit{i)} the $b$ and $c$ quark jets have $p_T>20~{\rm GeV}$,
\textit{ii)} the two leptons have $p_T>25~{\rm GeV}$,
\textit{iii)} all leptons and jets have pseudorapidity $|\eta|<2.4$, and
\textit{iv)} each jet is separated from the leptons by $\Delta R>0.4$.
For $90~{\rm GeV}\leq M_\phi\leq 160~{\rm GeV}$, the combined effect of these cuts is a minimum (maximum) signal acceptance ${\cal A}$ of 1.2\% (6.4\%) for $M_\phi=100$ (140) GeV.
From Ref.~\cite{CMS:2017wtu} we take the medium working points for the $b$ and $c$ tagging efficiencies of $\epsilon_b=68\%$\footnote{For $b$ tagging, this number corresponds to the ``DeepCSV''  tagger.}, $\epsilon_c=40\%$, respectively. 

We then calculate the event rate as 
\be
N_{\rm Sig.} =2 \sigma_{t\bar{t}}~{\cal A} ~\epsilon_b^2 \epsilon_c^2 ~{\cal B}(W\rightarrow \ell \nu)^2 {\cal B}(t\rightarrow \phi_t c)[1-{\cal B}(t\rightarrow \phi_t c)]~ {\cal L},
\ee
where the factor of 2 accounts for either the top or the anti-top quark undergoing the nonstandard decay, and ${\cal B}(W\rightarrow \ell\nu)\approx 0.25$ is the branching fraction for $W$ decays to an electron or muon, including those produced via $W\rightarrow\tau\nu$ with a subsequent leptonic $\tau$ decay.
After applying the selection cuts and efficiencies from~\cite{CMS:2020utv}, we find a maximum
fiducial cross section of $\sigma_{\rm fid}=N_{\rm Sig.}/{\cal L}\approx 0.121~{\rm pb}$ at
$M_{\phi}=150~{\rm GeV}$ and for a nonstandard top branching fraction of 50\%. To assess whether this is excluded, we derive
an upper limit on any additional contribution to the fiducial cross
section measured by CMS~\cite{CMS:2020utv} as
\begin{equation}
  \sigma_{\rm fid}^{\rm new} < \left(\sigma_{\rm fid}^{\rm obs}
  - \sigma_{\rm fid}^{\rm SM}\right) + 2\,\delta\sigma_{\rm fid} ~~,
\end{equation}
where $\sigma_{\rm fid}^{\rm obs}$ is the measured fiducial cross section,
$\sigma_{\rm fid}^{\rm SM}=0.187~{\rm pb}$ is the SM prediction at next-to-leading order, and
$\delta\sigma_{\rm fid}$ is the combined experimental, statistical and theoretical
uncertainty added in quadrature. This yields $\sigma_{\rm fid}^{\rm new} < 0.126~{\rm pb}$, so the
predicted signal
 is not excluded at the $2\sigma$ level.
  The ATLAS analysis~\cite{ATLAS:2024plw} used different event acceptance criteria, but likewise does not exclude the signal. 
We stress that both the CMS and ATLAS analyses considered here were designed for charm production from QCD radiation, and so are not optimized for this search. For example, both analyses required that the leptons be sufficiently separated from the jets; for our model, this cut is not optimal because the $\phi_t \rightarrow W^+ b \, \bar c$  decay produces jets and leptons that are closer together than when the charm jets are produced from QCD radiation. 
Moreover, the decay products of $\phi_t$ will of course reconstruct its invariant mass. While we consider a leptonically decaying $W$ (such that the $\phi_t$ mass cannot be completely reconstructed), the visible mass may prove to be an important handle to remove background.

In the right panel of Figure~\ref{fig:Nsig}, we show contours of constant event number as a function of the heavy scale $\Lambda_5$ and  $M_{\phi}$ during Run 3 of the LHC. 
For $M_{\phi}\lesssim 140~{\rm GeV}$, the event rate does not vary significantly, yielding about $10^3$ events for $\Lambda_5\approx 1~{\rm TeV}$. For larger $M_{\phi}$, the rate falls rapidly for two reasons: the phase space shrinks as $M_{\phi}$ approaches the top-quark mass (see Sec.~\ref{sec:model}), and the $c$ quark from the top decay becomes less likely to have a sufficiently large $p_T$ (see Figure~\ref{fig:Charmpt}).
This rate should be compared with the dominant background of top quark pair production in association with two additional charm jets, $t\bar{t}+c\bar{c}$, which, for the cuts we consider from the CMS analysis, has a fiducial cross section of $0.187~{\rm pb}$. This corresponds to roughly $6\times 10^4$ events in Run~3. The dashed red curve in Figure~\ref{fig:Nsig} is the event contour where the predicted number of signal events is equal to the statistical uncertainty on the background rate and serves as a sensitivity estimate.

While we have estimated the signal rate by adapting the selection cuts and tagging efficiencies of existing $t\bar t + c\bar c$ measurements, these analyses were not designed for the nonstandard decay $t \to \phi_t \, c$. 
A dedicated search that exploits the distinctive kinematics of the signal, such as the softer lepton and subleading charm jet from the $\phi_t$ decay, and the proximity of the $\phi_t$ decay products, may be sensitive to $\Lambda_5 \sim {\cal O} (1-3)~{\rm TeV}$.

\bigskip

%%%%%%%%%%%%%%%%%%%%%%%%%%%%%%%%%%%%%%%%%%%%%%%%%%%%%%%%%%%%
%%%%%%%%%%%%%%%%%%%%%%%%%%%%%%%%%%%%%%%%%%%%%%%%%%%%%%%%%%%%
\section{Conclusions}  
\label{sec:conclusions}    \setcounter{equation}{0}

Physics beyond the SM may lead to nonstandard top decays that can be searched for at colliders. The current 
limit on the branching fraction of nonstandard top decays, assuming hard to detect final states, is very weak, at around 26\% (see Figure~\ref{fig:BR} and the discussion in Section~\ref{sec:model}). Given that a few billion top quarks will be produced in the high-luminosity runs of the LHC, there are strong reasons for both theorists and experimental collaborations to investigate additional possible top decays.

Here we have studied a minimal renormalizable extension of the SM that allows nonstandard top decays involving a new scalar.
In this model, a gauge-singlet complex scalar, $\phi_t$, has an effective coupling to the top and charm quarks through a dimension-5 generated by a vectorlike quark $\chi$ that has Yukawa couplings to $\phi_t$ and the SM Higgs doublet.  
After electroweak symmetry breaking, the operator induces the  decay $t \to \phi_t \, c$.

While direct LHC searches for the vectorlike quark require $m_\chi \gtrsim 1.4$~TeV, it may still leave a substantial indirect imprint on top quark decays: the branching fraction $ {\cal B}(t \to \phi_t \, c)$ can be near the percent level.
Since more than $4 \times 10^8$ top quark pairs have been already produced in each of the ATLAS and CMS detectors, 
the above branching fraction corresponds to millions of events with nonstandard top decays.

The phenomenology of $\phi_t$ depends on its mass.
We focused on the range $M_W + m_b + m_c < M_\phi < m_t - m_c$, where $\phi_t$ decays through an off-shell top quark, $\phi_t \to t^* \, \bar c \to W^+ b \, \bar c$.
We computed the width of this 3-body decay analytically and found that it is highly sensitive to the $\phi_t$ mass. 
Throughout this mass range the  $\phi_t$  decay is prompt, and the competing 1-loop decay $\phi_t \to b \, \bar s$, which is suppressed by the charm and bottom masses, is negligible.
For a $\phi_t$ mass that is not far above $M_W$, the $\phi_t$ decay products from $\phi_t\rightarrow W b \bar c$ carry relatively little momentum, which shapes the collider signatures of the nonstandard top decay; similarly, at the other limit as $M_\phi$ approaches the top mass, the charm quark from $t\rightarrow \phi_t \, c$ is produced with relatively smaller momenta, which plays an important role for charm tagging. 

At the LHC, the nonstandard top decay leads to the final state $W^+ W^- b \, \bar b \, c \, \bar c$, which also arises in the SM from $t\bar t$ production in association with a $c\bar c$ pair.
Existing ATLAS and CMS measurements of $t\bar t + c\bar c$ production do not yet constrain our model, but they provide a useful baseline: adopting the selection cuts and tagging efficiencies of the CMS measurement, we estimate that Run~3 data could be sensitive to $m_\chi$ up to about 2~TeV, assuming that its product of Yukawa couplings is of order one.
These measurements were designed for charm quarks produced by QCD radiation, and their selection is not well suited for the decay products of $\phi_t$.
A dedicated search, for example one with lower charm-tagging thresholds, relaxed lepton--jet separation, or reconstruction of the $W^+ b \, \bar c$ visible invariant mass, could substantially improve the sensitivity to $m_\chi$ up to a few TeV. 

There is an interesting complementarity between searches for nonstandard top decays and searches for heavy new particles that may be responsible for the top decays. In the model discussed here, the vectorlike quark can be singly produced (in association with a $\phi_t$) at colliders. Dedicated searches for this process at the LHC would likely be sensitive to $m_\chi$ even above 2 TeV, almost independently of the product of Yukawa couplings. The vectorlike quark may decay mostly as $\chi \to \phi_t \, c$, depending on the ratio of its Yukawa couplings. The ensuing LHC signal is $pp \to \chi \phi_t \, j \to  3c + b \, \bar b \, W W j$ with one of $cbW$ systems being highly boosted. 
Looking beyond the LHC, nonstandard top quark decays are well-motivated targets for searches at an  FCC-ee \cite{FCC:2018evy} run above the $t\bar t$ threshold, while vectorlike quarks like the one  analyzed here would be well probed at the FCC-hh \cite{FCC:2018vvp} or at a high-energy muon collider \cite{InternationalMuonCollider:2025sys}.

\bigskip\bigskip\bigskip

%%%%%%%%%%%%%%%%%%

{\bf Acknowledgments:} \ Fermilab is administered by Fermi Forward Discovery Group, LLC under Contract No. 89243024CSC000002 with the U.S. Department of Energy, Office of Science, Office of High Energy Physics.
AR is supported by Stanford University under Contract No. DE-AC02-76SF00515 with the U.S. Department of Energy, Office of Science, Office of High Energy Physics; by the National Science Foundation under Grant No. PHY-2310429; by the Simons Foundation through Investigator Award No. 824870; and by the John Templeton Foundation under Award No. 63595.

\smallskip
\smallskip

%%%%%%%%%%%%%%%%%%%%%%%%%%%%%%%%%%%%%%%%%%%%%%%%%%%%%%%%%%%%
%%%%%%%%%%%%%%%%%%%%%%%%%%%%%%%%%%%%%%%%%%%%%%%%%%%%%%%%%%%%

\end{document}